\documentstyle[12pt]{article}
 \makeatletter \oddsidemargin  -.1in \evensidemargin -.1in
\newcommand{\singlespacing}{\let\CS=\@currsize\renewcommand{\baselinestretch}{1}\tiny\CS}
\newcommand{\oneandahalfspacing}{\let\CS=\@currsize\renewcommand{\baselinestretch}{1.25}\tiny\CS}
\newcommand{\doublespacing}{\let\CS=\@currsize\renewcommand{\baselinestretch}{1.35}\tiny\CS}

\newtheorem{rule-def}[theorem]{Rule}

\RequirePackage[dvips]{graphicx} \textheight 21.5cm
\begin{document}
\title{\bf Generalized Magnetothermoelastic Interaction for a Rotating Half Space}
\author{\small B. Das,  \thanks{Corresponding Author:
bappa.das1@gmail.com (B. Das)} $^{}$
~~ ~~\small S. Chakraborty $^{\dag}$ ~and ~A. Lahiri \thanks{Email Address: csanjukta1977@gmail.com (S.  Chakraborty), lahiriabhijit2000@yahoo.com (A. Lahiri)
} \\
\it {$^*$}Department of Mathematics,\\
\it Ramakrishna Mission Vidyamandira,\\\it Belur Math, Howrah - 711202.\\
\it {$^\dag$}Department of Mathematics,\\
\it Jadavpur University,Kolkata - 700032. }
\date{}
\maketitle \noindent \doublespacing
\noindent {\bf Abstract -}
A generalized magnetothermoelasticity, in the context of Lord-Shulman theory, is employed to investigate the interaction of a homogeneous and isotropic perfect conducting half space with rotation. The Laplace transform for time variable is used to formulate a vector-matrix differential equation which is then solved by eigenvalue method. The continuous solution of displacement component while the discontinuous solutions of stress components, temperature distribution, induced magnetic and electric field have been analyzed in an approximate manner using assymptotic expansion for small time. The graphical representations also prove this continuity and discontinuity of the solutions.

\section{\large{\bf{ {Introduction}}}}
The classical theory of uncoupled thermoelasticity predicts two phenomenon not suitable with experimental results such as (i) the heat conduction equation of this theory does not contain any elastic terms but the fact that the elastic changes produce thermal effect and (ii) the heat conduction equation is of parabolic type whereas the equation of motion is of hyperbolic type, predicting infinite speed of propagation for predominantly thermal disturbances.
 The heat conduction equation is based on classical Fourier's law $\overline{q}=-k\overline{\nabla}T$ relates heat flux $\overline{q}$ to the temperature gradient $T$, $k$ is the thermal conduction in Ignaczak and Ostoja-Starzewski\cite{R1}.
\paragraph{}  There are mainly two different models of generalized thermoelasticity (i) Lord and Shulman \cite{R2}(L-S model) which modifies the Fourier's law of heat conduction introducing one relaxation time parameter. This L-S model is also known as extended thermoelasticity (ETE) and (ii) the second generalization is given by Green and Lindsay \cite{R3}(G-L model) with two relaxation time parameters which is based on a generalized inequality of thermodynamics. This G-L model is also known as temperature rate dependent thermoelasticity (TRDTE). In G-L model, the equations of both motion and heat conduction are hyperbolic and the equation of motion is modified and differ from Coupled thermoelasticity (CTE).
 
\paragraph{} Sherief and Helmy \cite{R4} studied a one dimensional and two dimensional generalized magnetothermoelastic problems, respectively. Tupholme \cite{R5}, Ezzat and Bary \cite{R6} and Sarkar and Lahiri \cite{R7} also studied many problems related to this field.
\paragraph{} This paper concerns the problem of a generalized magnetothermoelastic interaction of a homogeneous, isotropic perfectly conducting half-space with  rotation. The Laplace transform is used for time variable to formulate the vector-matrix differential equation which is then solved by eigenvalue techniques. The inverse of Laplace transform is obtained in an approximate manner using assymptotic expansion for small time. Finally, the numerical computations of displacement, stresses, temperature, induced magnetic and electric fields have been done and analyzed graphically.

\section{\large{\bf{{\small{\bf {Basic Equations }}}}}}

The field equations and electromagnetic quantities satisfy Maxwell's equations for generalized thermo-elastic body may be written as-
\begin{eqnarray}
\textrm{curl} ~{\bf{h}} = {\bf{J}} + \varepsilon_0\frac{\partial {\bf{E}}}{\partial t}~,~
\textrm{curl} ~{\bf{E}} = - \mu_0\frac{\partial {\bf{h}}}{\partial t}\nonumber\\
\textrm{div} ~ {\bf{h}} = 0,~~~~~~~~{\bf{E}} = -\mu_0(\dot{\bf{u}}\times \bf{H})\nonumber\\
{\bf{B}} = \mu_0({\bf{H}}+ {\bf{h}}),~~~~~~~~{\bf{D}} = \varepsilon_0{\bf{E}}
\end{eqnarray}
And the equation for the Lorentz force is
\begin{eqnarray}
{\bf{F}} = {\bf{J}}\times{\bf{B}}= \mu_0({\bf{J}}\times \bf{H})
\end{eqnarray}
In the absence of body force and internal heat source, the generalized electro-magneto-thermo-elastic equations of motion in the context of L-S model in account of the Lorentz force is given by
\begin{eqnarray}
\sigma_{ji,j} + \mu_0({\bf{J}}\times \bf{H})_i = \rho[\ddot{u}_i+({\bf{\Omega}}\times({{\bf{\Omega}}\times{{\bf u}}}))_i+
({2{{\bf {\Omega}}\times\dot{{{\bf u}}}}})_i]
\end{eqnarray}

The constitutive stress components are given by the Hooke-Duhamel-Neumann law
\begin{eqnarray}
\sigma_{ij} = 2\mu e_{ij} + \lambda e \delta_{ij} - \gamma(T-T_0)\delta_{ij}
\end{eqnarray}

The strain tensors are given by
\begin{eqnarray}
e_{ij} = \frac{1}{2}(u_{i,j} + u_{j,i})
\end{eqnarray}
Now, the heat conduction equation is
\begin{eqnarray}
kT_{,ii}  = \rho c_E(\dot{T}+\tau_0\ddot{T})+
\gamma T_0(\dot{e}_{kk}+\tau_0\ddot{e}_{kk})
\end{eqnarray}
 where, $\lambda,\mu $ are Lam$\grave{e}$  constants,
$T $ is the absolute temperature,
$T_0$ is the reference temperature chosen such that
$|\frac{T-T_0}{T_0}|<<1$,
$\rho$ is the density of the medium,
$\gamma$ is the material Constant = $({3\lambda+2\mu}){\alpha_T}$,
$\alpha_T$ is also the coefficient of linear thermal expansion,
$t$ is the time variable,
$C_E$ is the specific heat at constant strain,
$\tau _0$ is the thermal relaxation time parameter,
${\bf{u}}$ is the displacement vector,
${\bf{h}} $ is the induced magnetic field vector,
${\bf{J}}$ is the electric current density vector,
${\bf{B}}$ is the magnetic induction vector,
${\bf{D}}$ is the electric induction vector,
$\varepsilon_0$ is the electric permeability,
$\mu_0$ is the magnetic permeability,
$\sigma_0$ is the electric conductivity,
$k$ is coefficient of thermal conductivity,
${\bf{\Omega}}$ is the rotational vector applied along z-axis,
$e = \textrm{div} {\bf{u}}$ is the cubical dilatation, $\delta_{ij}$ is Kronecker's delta tensor and $e_{ij}$ is the strain tensor.\\

\section{\large{\bf{{\small{\bf {Formulation of the Problem}}}}}}

We now consider the problem of  a perfectly homogeneous, isotropic thermoelastic half-space ($x\geq0$) under a magnetic field with constant intensity $H_0$ acts tangentially to the bounding plane and an induced electric field ${\bf{E}}$. We also assume that both ${\bf{h}}$ and ${\bf{E}}$ are small in magnitude in accordance with the assumption of the linear theory of thermoelasticity. In this case, we consider $u_x = u(x,t)$, $u_y = u_z=0$, $e_{xx} = \frac{\partial u}{\partial x}$  and the component of Lorentz force along $x$-direction is

$F_x=-\mu_0H_0(-H_0\frac{\partial^2 u}{\partial x^2}+ \varepsilon_0\mu_0H_0\frac{\partial^2 u}{\partial t^2})$.\\

Equation (1) gives
\begin{eqnarray}
({{J}},{{E}},{{h}}) = (\frac{\partial h}{\partial x}- \varepsilon_0 \frac{\partial E}{\partial t} ~,~-\mu_0H_0\frac{\partial u}{\partial t}~,~-H_0\frac{\partial u}{\partial x})
\end{eqnarray}

Using (7) in (3), we get the equation of motion as given below
\begin{eqnarray}
(\lambda + 2\mu+\mu_0H^2_0)\frac{\partial^{2} u}{\partial x^2}- \gamma\frac{\partial T}{\partial x}
 -(\varepsilon_0\mu^2_0H^2_0+ \rho)\frac{\partial^{2} u}{\partial t^2}+\rho \Omega^2 u=0
\end{eqnarray}
The heat conduction equation is
\begin{eqnarray}
k\frac{\partial^{2} T}{\partial x^2}=(\frac{\partial}{\partial t} + \tau_0\frac{\partial^{2} }{\partial t^2})(\rho c_E T + \gamma T_0\frac{\partial u}{\partial x})
\end{eqnarray}
The stress components are
\begin{eqnarray}
 (\sigma_{xx},\sigma_{yy}=\sigma_{zz})=((\lambda+2\mu)\frac{\partial u}{\partial x} - \gamma(T - T_0),\lambda\frac{\partial u}{\partial x} - \gamma(T - T_0))
 \end{eqnarray}
To obtain the non-dimensional form of equations (8) - (10), we use the following non-dimensional variables as follows
\begin{eqnarray}
(x^*,u^*)=c_0\eta (x,u)~,~(t^*,\tau_0^*)=c_0^2\eta (t,\tau_0)~,~\nonumber\\
(\theta^*,\sigma^*_{ij})=\frac{1}{(\lambda+2\mu)}(\gamma(T - T_0),\sigma_{ij})~, ~(h^*,E^*)=\frac{1}{\mu_0H_{0}c_0}(\mu_0c_0h, E)
\end{eqnarray}
Dropping the asterisks for convenience, we obtain the non-dimensional form of equation of motion, heat conduction equation and stress components respectively, as-
\begin{eqnarray}
\beta^2\frac{\partial^{2} u}{\partial x^2}- \beta_0^2\frac{\partial \theta}{\partial x}-\alpha_0\frac{\partial^{2} u}{\partial t^2}+\varepsilon_1u=0
\end{eqnarray}
\begin{eqnarray}
\frac{\partial^{2} \theta}{\partial x^2}= (\frac {\partial}{\partial t} + \tau_0\frac{\partial^2 }{\partial t^2})(\theta+\varepsilon_2 \frac{\partial u}{\partial x})
\end{eqnarray}

\begin{eqnarray}
(\sigma_{xx},\sigma_{yy}) = (\frac{\partial u}{\partial x}-\theta~,~(1-\frac{2}{\beta_0^2})\frac{\partial u}{\partial x}-\theta)
\end{eqnarray}
where,
 $\beta^2$=$\frac{\lambda+2\mu}{\mu}+\frac{\mu_0H^2_0}{\mu}$~;~$\beta^2_0$=$\frac{\lambda+2\mu}{\mu}$~;~$\alpha_0$ = $\frac{(\varepsilon_0\mu_0^2H^2_0+\rho)c_0^2}{\mu}$~;~$\varepsilon_1$=$\frac{\rho \Omega^2}{c_0^2\eta^2\mu}$ ~;~
$\varepsilon_2$=$\frac{\gamma^2 T_0}{C_E\rho^2c_0^2}$~;~$c_0^2$ = $\frac{\lambda+2\mu}{\rho}$~;~
$C_E$=$\frac{k\eta}{\rho}$.\\

\section{\large{\bf{{Solution Procedure : Eigenvalue Technique}}}}
We now apply the Laplace transform defined by
\begin{eqnarray}
L[f(x,t)]=\bar{f}(p)={\int^{\infty}_{0}f(x,t)exp(-pt)dt}~~,~~Re(p)>0
\end{eqnarray}
to the equations (12)-(14), we get
\begin{eqnarray}
\frac{d^{2} \bar{u}}{dx^2} = (\frac{\alpha_0p^2}{\beta^2}-\frac{\varepsilon_1}{\beta^2})\bar{u}+
\frac{\beta_0^2}{\beta^2}\frac{d \bar {\theta}}{dx} \\
\frac{d^{2} \bar{\theta}}{dx^2} = (p+\tau_0p^2)\bar{\theta}+\varepsilon_2(p+\tau_0p^2)
\frac{d \bar {u}}{dx}
\end{eqnarray}

\begin{eqnarray}
(\bar{\sigma}_{xx},\bar{\sigma}_{yy})=(\frac{d \bar {u}}{dx}-\bar{\theta},(1-\frac{2}{\beta^2_0})\frac{d\bar{u}}{dx}-\bar{\theta})
\end{eqnarray}
Initially i.e., at time $t=0$, the displacement component and temperature along with their derivatives with respect to $t$ are zero and maintained at the reference temperature $T_0$, so the following initial conditions hold.
\begin{eqnarray}
u(x,0)=\frac{\partial u(x,0)}{\partial
t}=0~;~~\frac{\partial T(x,0)}{\partial
t}=0~~and~~ T(x,0)=T_0
\end{eqnarray}
As in Sarkar and Lahiri \cite{R7} also in Das and Lahiri \cite{R8}, equations(16) and (17) can be written as-
\begin{eqnarray}
D\underline{V}(x,p) = \underline{A}(p)~\underline{V}(x,p)~~;~~D\equiv\frac{d}{dx}
\end{eqnarray}
Where,

$\underline{V}(x,p)=\left[{\bar{u}}~~~{\bar{\theta}}~~~D{\bar{u}}
~~~D{\bar{\theta}}\right]^T$,

$\underline A(p)=\left[\begin{array}{clcr}
 L_{11} & L_{12} \\
 L_{21} & L_{22}
  \end{array} \right]$, $L_{11}$ and $L_{12}$ are null and identity matrix of order 2 respectively and $L_{21}$ and $L_{22}$ are given in the Appendix-I.\\

For the solution of the vector-matrix differential equation(20),
we now apply the method of eigenvalue method as in Sarkar and Lahiri \cite{R7} also in Das and Lahiri \cite{R8}. The characteristic equation of the matrix $\underline A(p)$ can be written as
\begin{eqnarray}
k^4 - ak^2 + b= 0
\end{eqnarray}
Where
\begin{eqnarray}
a=C_{34}C_{43}+C_{42}+C_{31}~~~~~~~~~~~~~~~~~~~~~~~~~\nonumber\\=a_3\varepsilon_2(p+\tau_0 p^2)+(p+\tau_0 p^2)+(a_1p^2-a_2)~;\nonumber\\b=C_{31}C_{42}=(a_1p^2-a_2)(p+\tau_0p^2)~~~~~~~~~~~~~
\end{eqnarray}
The roots with positive real parts of the characteristic equation (21) are also the eigenvalues of the matrix $\underline A$ which are of the form $k=k_{1,2}$
\begin{eqnarray}
k_{1,2}=[\frac{1}{2}\{a\pm(a^2-4b)^{\frac{1}{2}}\}]^{\frac{1}{2}}
\end{eqnarray}
The eigenvectors $\underline X$ of the matrix $\underline A$ corresponding to the eigenvalues $k$ can be
calculated as-
\begin{eqnarray}
\underline {X}=\left[x_1~~~x_2~~~x_3
~~~x_4\right]^T~~~~~~~~~~~~~~~~~~~~~~~~~~~~~~~~\nonumber\\=\left[ka_3~~~(k^2-a_1p^2+a_2)~~~k^2a_3
~~~k(k^2-a_1p^2+a_2)\right]^T
\end{eqnarray}
Considering the regularity condition at infinity, as in Sarkar and Lahiri \cite{R7} also in Das and Lahiri \cite{R8}, the general solution of equation (20) can be written as-
\begin{eqnarray}
\underline {V}(x)(x,p)= \sum_{i=1}^{2}{A_i}x_{i}e^{-k_ix}
\end{eqnarray}
 where, $A_i$'s $(i=1,2)$ are the arbitrary parameters determined from the boundary conditions.\\
Thus the components of $\underline {V}(x,p)$ i.e., ${\bar{u}}$ and ${\bar{\theta}}$ are calculated in Laplace transform domain such as-
\begin{eqnarray}
[\bar u,\bar \theta]=\sum_{i=1}^{2}{A_i}[-a_3k_i,(k_i^2-a_1p^2+a_2)]e^{-k_i x}
\end{eqnarray}
With the help of equations (11) and (15), equation (7) becomes
\begin{eqnarray}
(\bar h,\bar E)=a_3\sum_{i=1}^{2}{A_i}k_i(-k_i,p)e^{-k_i x}
\end{eqnarray}
From equation (18), we get the stress components as-
\begin{eqnarray}
\bar{\sigma}_{xx}=\sum_{i=1}^{2}A_i[k_i^2a_3-(k_i^2-a_1p^2+a_2)]e^{-k_i x}\nonumber\\
\bar{\sigma}_{yy}=\sum_{i=1}^{2}A_i[(1-\frac{2}{\beta_0^2})k_i^2a_3-(k_i^2-a_1p^2+a_2)]e^{-k_i x}
\end{eqnarray}
\section{\large{\bf{Boundary Conditions}}}
We now consider a homogeneous elastic medium of perfectly conducting half-space occupying the region $x\geq 0$ with quiescent initial state.\\
The non-dimensional boundary conditions at $x=0$ are\\
(i) Thermal shock :
\begin{eqnarray}
              \theta(0,t)=\theta_0H(t)
\end{eqnarray}

(ii) Mechanical boundary condition :
\begin{eqnarray}
              \sigma_{xx}(0,t)+T_{xx}(0,t)-T_{xx}^0(0,t)=0
\end{eqnarray}
where, $\theta_0$ is a constant temperature and $H(t)$ is also Heaviside unit step function.\\
Since the transverse components of the vectors $\bf{E}$ and $\bf{h}$ are continuous across the boundary plane i.e., $E(0,t)=E^0(0,t)$ and $h(0,t)=h^0(0,t), t>0$ where $E^0$ and $h^0$ are the components of the induced electric and magnetic field in free-space and the relative permeabilities are very nearly unity, it follows that $T_{xx}(0,t)=T_{xx}^0(0,t)$ and equation (30) reduces to
\begin{eqnarray}
              \sigma_{xx}(0,t)=0
\end{eqnarray}
where $T_{xx}(0,t)$ is the Maxwell stress tensor and $T_{xx}^0(0,t)$ is the Maxwell stress tensor in a vacuum.\\
Equations (29) and (31) become in Laplace transform domain
\begin{eqnarray}
              [\bar{\theta}(0,p),\bar{\sigma}_{xx}(0,p)]=\frac{\theta_0}{p}[1,0]
\end{eqnarray}
With the help of equations (32), (26) and (28), we get
\begin{eqnarray}
              [A_1,A_2]=\frac{\theta_0}{pa_3(a_1p^2-a_2)(k_1^2-k_2^2)}[\beta_{22},\beta_{11}]
\end{eqnarray}
where, $\beta_{11}=k_1^2a_3-k_1^2+a_1p^2-a_2$ and $\beta_{22}=k_2^2a_3-k_2^2+a_1p^2-a_2$.

\section{\large{\bf{Inversion of Laplace Transform}}}
We now calculate the inverse Laplace transforms for the case of small values of time (large values of p). Denoting $u=p^{-1}$, we have
\begin{eqnarray}
              k_i=u^{-1}[f_i(u)]^{\frac{1}{2}},~~i=1,2
\end{eqnarray}
where,
\begin{eqnarray}
              f_i(u)=\frac{1}{2}[\{(u+\tau_0)(a_3\varepsilon_2+1)+(a_1-a_2u^2)\}]\pm\nonumber\\
              \frac{1}{2}[\{(u+\tau_0)(a_3\varepsilon_2+1)+(a_1-a_2u^2)\}^2-4(a_1-a_2u^2)(u+\tau_0)]^{\frac{1}{2}}
\end{eqnarray}
Expanding $f_i(u)$ in the Maclaurin series of which the first three terms are retained, we get
\begin{eqnarray}
f_i(u) = \sum_{j=0}^{2}a_{ij}u^{j}, i=1,2
\end{eqnarray}
where, the values of $a_{ij}$ are given in the Appendix-I. \\

Next, we expand the expressions $[f_i(u)]^{\frac{1}{2}}$ in a Maclaurin series and retaining the first three terms, we finally obtain the expressions for $k_i$ in the form
\begin{eqnarray}
k_i = u^{-1}\sum_{j=0}^{2}b_{ij}u^{j}, i=1,2
\end{eqnarray}
where,
$b_{i0}=\sqrt{a_{i0}}$, $b_{i1}=\frac{a_{i1}}{2\sqrt{a_{i0}}}$,$b_{i2}=\frac{(4a_{i0}a_{i2}-a_{i1}^2)}{8a_{i0}^{\frac{3}{2}}}$.\\
From equation (37), we get the value of $k_i$'s for $i=1,2$ respectively and hence
\begin{eqnarray}
\frac{1}{k_1^2-k_2^2} = (b_{10}^2-b_{20}^2)u^{2}-2(b_{10}b_{11}-b_{20}b_{21})u^3-\{(b_{11}^2-b_{21}^2)+2(b_{10}b_{12}
-b_{20}b_{22})\}u^4\nonumber\\
-2(b_{11}b_{12}-b_{21}b_{22})u^5-(b_{12}^2-b_{22}^2)u^{6}~~~~~~~~~~~~~~~~~~~~~~~~~~~~~~~~~~~~~~~~~~~~~~~~~~~
\end{eqnarray}

The closed form solutions of the displacement, temperature, induced magnetic and electric fields also the stresses for isotropic half space under effect of rotation have been obtained in L-S theory and these are
\begin{eqnarray}
\bar u = -\frac{\theta_0}{a_1}[e^{-b_{11}x}\sum_{j=1}^{3}\frac{b_j}{p^{j+1}}e^{-b_{10}xp}e^{-x\frac{b_{12}}{p}}
-e^{-b_{21}x}\sum_{j=1}^{3}\frac{c_j}{p^{j+1}}e^{-b_{20}xp}e^{-x\frac{b_{22}}{p}}]
\end{eqnarray}
\begin{eqnarray}
\bar \theta = \frac{\theta_0}{a_1a_3}[e^{-b_{11}x}\sum_{j=0}^{3}\frac{d_j}{p^{j+1}}e^{-b_{10}xp}e^{-x\frac{b_{12}}{p}}
-e^{-b_{21}x}\sum_{j=0}^{3}\frac{e_j}{p^{j+1}}e^{-b_{20}xp}e^{-x\frac{b_{22}}{p}}]
\end{eqnarray}
\begin{eqnarray}
\bar h = -\frac{\theta_0}{a_1}[e^{-b_{11}x}\sum_{j=0}^{3}\frac{f_j}{p^{j+1}}e^{-b_{10}xp}e^{-x\frac{b_{12}}{p}}
-e^{-b_{21}x}\sum_{j=0}^{3}\frac{g_j}{p^{j+1}}e^{-b_{20}xp}e^{-x\frac{b_{22}}{p}}]
\end{eqnarray}
\begin{eqnarray}
\bar E = \frac{\theta_0}{a_1}[e^{-b_{11}x}\sum_{j=0}^{3}\frac{h_j}{p^{j+1}}e^{-b_{10}xp}e^{-x\frac{b_{12}}{p}}
-e^{-b_{21}x}\sum_{j=0}^{3}\frac{k_j}{p^{j+1}}e^{-b_{20}xp}e^{-x\frac{b_{22}}{p}}]
\end{eqnarray}
\begin{eqnarray}
\bar {\sigma}_{xx} = \frac{\theta_0}{a_1a_3}[e^{-b_{11}x}\sum_{j=0}^{3}\frac{m_j}{p^{j+1}}e^{-b_{10}xp}e^{-x\frac{b_{12}}{p}}
-e^{-b_{21}x}\sum_{j=0}^{3}\frac{n_j}{p^{j+1}}e^{-b_{20}xp}e^{-x\frac{b_{22}}{p}}]
\end{eqnarray}
\begin{eqnarray}
\bar {\sigma}_{yy} = \frac{\theta_0}{a_1a_3}[e^{-b_{11}x}\sum_{j=0}^{3}\frac{r_j}{p^{j+1}}e^{-b_{10}xp}e^{-x\frac{b_{12}}{p}}
-e^{-b_{21}x}\sum_{j=0}^{3}\frac{s_j}{p^{j+1}}e^{-b_{20}xp}e^{-x\frac{b_{22}}{p}}]
\end{eqnarray}
To invert the Laplace transform, we now use the basic theorem and formulas for the Laplace transforms, namely Oberhetinger and Badii\cite{R9} which are given in the Appendix-II, taking $J_{\nu}$ and $I_{\nu}$ are the Bessel and the modified Bessel functions of order $\nu$ of the first kind respectively and in the absence of the applied magnetic field, we always have $b_{12}>0$ and $b_{22}<0$.\\ Equations (39)-(44) transforms to-
\begin{eqnarray}
u = -\frac{\theta_0}{a_1}[e^{-b_{11}x}H(t-b_{10}x)\sum_{j=1}^{3}b_j(\frac{t-b_{10}x}{b_{12}x})^{\frac{j}{2}}J_j(z_1)
\nonumber\\-e^{-b_{21}x}H(t-b_{20}x)\sum_{j=1}^{3}c_j(\frac{t-b_{20}x}{-b_{22}x})^{\frac{j}{2}}I_j(z_2)]
\end{eqnarray}
\begin{eqnarray}
\theta = \frac{\theta_0}{a_1a_3}[e^{-b_{11}x}H(t-b_{10}x)\sum_{j=0}^{3}d_j(\frac{t-b_{10}x}{b_{12}x})^{\frac{j}{2}}J_j(z_1)
\nonumber\\-e^{-b_{21}x}H(t-b_{20}x)\sum_{j=0}^{3}e_j(\frac{t-b_{20}x}{-b_{22}x})^{\frac{j}{2}}I_j(z_2)]
\end{eqnarray}
\begin{eqnarray}
 h= -\frac{\theta_0}{a_1}[e^{-b_{11}x}H(t-b_{10}x)\sum_{j=0}^{3}f_j(\frac{t-b_{10}x}{b_{12}x})^{\frac{j}{2}}J_j(z_1)
\nonumber\\-e^{-b_{21}x}H(t-b_{20}x)\sum_{j=0}^{3}g_j(\frac{t-b_{20}x}{-b_{22}x})^{\frac{j}{2}}I_j(z_2)]
\end{eqnarray}
\begin{eqnarray}
 E= \frac{\theta_0}{a_1}[e^{-b_{11}x}H(t-b_{10}x)\sum_{j=0}^{3}h_j(\frac{t-b_{10}x}{b_{12}x})^{\frac{j}{2}}J_j(z_1)
\nonumber\\-e^{-b_{21}x}H(t-b_{20}x)\sum_{j=0}^{3}k_j(\frac{t-b_{20}x}{-b_{22}x})^{\frac{j}{2}}I_j(z_2)]
\end{eqnarray}
\begin{eqnarray}
 {\sigma}_{xx}= \frac{\theta_0}{a_1a_3}[e^{-b_{11}x}H(t-b_{10}x)\sum_{j=0}^{3}m_j(\frac{t-b_{10}x}{b_{12}x})^{\frac{j}{2}}J_j(z_1)
\nonumber\\-e^{-b_{21}x}H(t-b_{20}x)\sum_{j=0}^{3}n_j(\frac{t-b_{20}x}{-b_{22}x})^{\frac{j}{2}}I_j(z_2)]
\end{eqnarray}
\begin{eqnarray}
 {\sigma}_{yy}= \frac{\theta_0}{a_1a_3}[e^{-b_{11}x}H(t-b_{10}x)\sum_{j=0}^{3}r_j(\frac{t-b_{10}x}{b_{12}x})^{\frac{j}{2}}J_j(z_1)
\nonumber\\-e^{-b_{21}x}H(t-b_{20}x)\sum_{j=0}^{3}s_j(\frac{t-b_{20}x}{-b_{22}x})^{\frac{j}{2}}I_j(z_2)]
\end{eqnarray}

where, the values of the $b_j$, $c_j$, $d_j$, $e_j$, $f_j$, $g_j$, $h_j$, $k_j$, $m_j$, $n_j$, $r_j$, $s_j$, $z_1$ and $z_2$ are given in Appendix-III.\\

\section{\large{\bf{Numerical Results and Discussions}}}
With an aim to illustrate the problem, we will present some numerical results. For this purpose, numerical computation is carried out for following physical parameters in SI units.\\
$\lambda  = 7.76\times10^{10}$~~;~~$\mu  = 3.86\times10^{10}$~~;~~
$\mu_0 = 4\pi\times10^{-7}$~~;~~$\alpha_t  = 1.78\times10^{-5}$~~;~~$H_0=\frac{10^{7}}{4\pi}$~~;
$\varepsilon_0=0.0168$~~;~~$T_0 = 293$~~~;~~~$c_E=383.1$~~~;~~~
$\tau=0.02$~~~;~~~$\sigma_0=5.7\times10^{7}$~~~;~~~$k=386$~~~;
$\rho = 8954$~~;~~$t = 0.3$~~;~~$\eta = 1$~~;~~$\gamma = 1$\\

1.~~From the equations (46), (47), (48), (49) and (50), we have come to the following conclusion-\\
            (i)~~~~~~~For fixed values of time$(t)$, temperature $(\theta)$, induced magnetic$(h)$ and electric$(E)$ field also stresses $(\sigma_{xx}$ and $\sigma_{yy})$ are continuous function for $0\leq x<\infty$ except at the points $x_1=\frac{t}{b_{10}}$ and $x_2=\frac{t}{b_{20}}$.\\
            (ii)~~~~~~~Two stresses $(\sigma_{xx})$ and $(\sigma_{yy})$ are similar of their manner.\\
             (iii)~~~~~~~Table-1 shows the expressions of jumps of discontinuities for $\theta$, $h$, $E$, $\sigma_{xx}$ and $\sigma_{yy}$ at the points $x_1$ and $x_2$.
\begin{center}
 Table~-~1\\
 $ $
$\begin{tabular}{|l|r|r|r|r|r|r|r|}   \hline
\emph{ }&
\emph{$\theta$} & \emph{$h$}  & \emph{$E$} & \emph{$\sigma_{xx}$} & \emph{$\sigma_{yy}$} \\  \hline
\emph{$x_1$}&
 \emph{$\frac{\theta_0}{a_1a_3}d_0e^{-\frac{b_{11}t}{b_{10}}}$} & \emph{$-\frac{\theta_0}{a_1}f_0e^{-\frac{b_{11}t}{b_{10}}}$} & \emph{$\frac{\theta_0}{a_1}h_0e^{-\frac{b_{11}t}{b_{10}}}$}& \emph{$-\frac{\theta_0}{a_1a_3}m_0e^{-\frac{b_{11}t}{b_{10}}}$} & \emph{$-\frac{\theta_0}{a_1a_3}r_0e^{-\frac{b_{11}t}{b_{10}}}$} \\    \hline
\emph{$x_2$} &
\emph{$-\frac{\theta_0}{a_1a_3}e_0e^{-\frac{b_{21}t}{b_{20}}}$} & \emph{$\frac{\theta_0}{a_1}g_0e^{-\frac{b_{21}t}{b_{20}}}$} & \emph{$-\frac{\theta_0}{a_1}k_0e^{-\frac{b_{21}t}{b_{20}}}$}& \emph{$\frac{\theta_0}{a_1a_3}n_0e^{-\frac{b_{21}t}{b_{20}}}$}& \emph{$\frac{\theta_0}{a_1a_3}s_0e^{-\frac{b_{21}t}{b_{20}}}$} \\ \hline
\end{tabular}$\\
\end{center}
              (iv)~~~~~~~Table-2 shows that the numerical values of the jump of discontinuities for fixed values of
               times $t_i$, where $i=1(1)5$.
\vspace*{0.5cm}
\begin{center}
 Table~-~2\\
 $ $
$\begin{tabular}{|l|r|r|}   \hline
\emph{$At~ the~ time$ }&
\emph{$x_{1}$} & \emph{$x_2$}  \\  \hline
\emph{$t_1=0.04$}&
 \emph{$0.037$} & \emph{$0.4$} \\    \hline
\emph{$t_2=0.2$} &
\emph{$0.186$} & \emph{$2$}  \\ \hline
\emph{$t_3=0.4$} &
\emph{$0.373$} & \emph{$4$} \\ \hline
\emph{$t_4=0.05$} &
\emph{$0.047$} & \emph{$0.5$} \\ \hline
\emph{$t_5=0.07$} &
\emph{$0.066$} & \emph{$0.7$} \\ \hline
\end{tabular}$\\
\end{center}
We have drawn several graphs for different values of
the space and time variable and conclude that-\\

\textheight 22.0cm 
\begin{minipage}{1.0\textwidth}
\vspace*{1.5cm}
\includegraphics[width=4.5in,height=2.5in ]{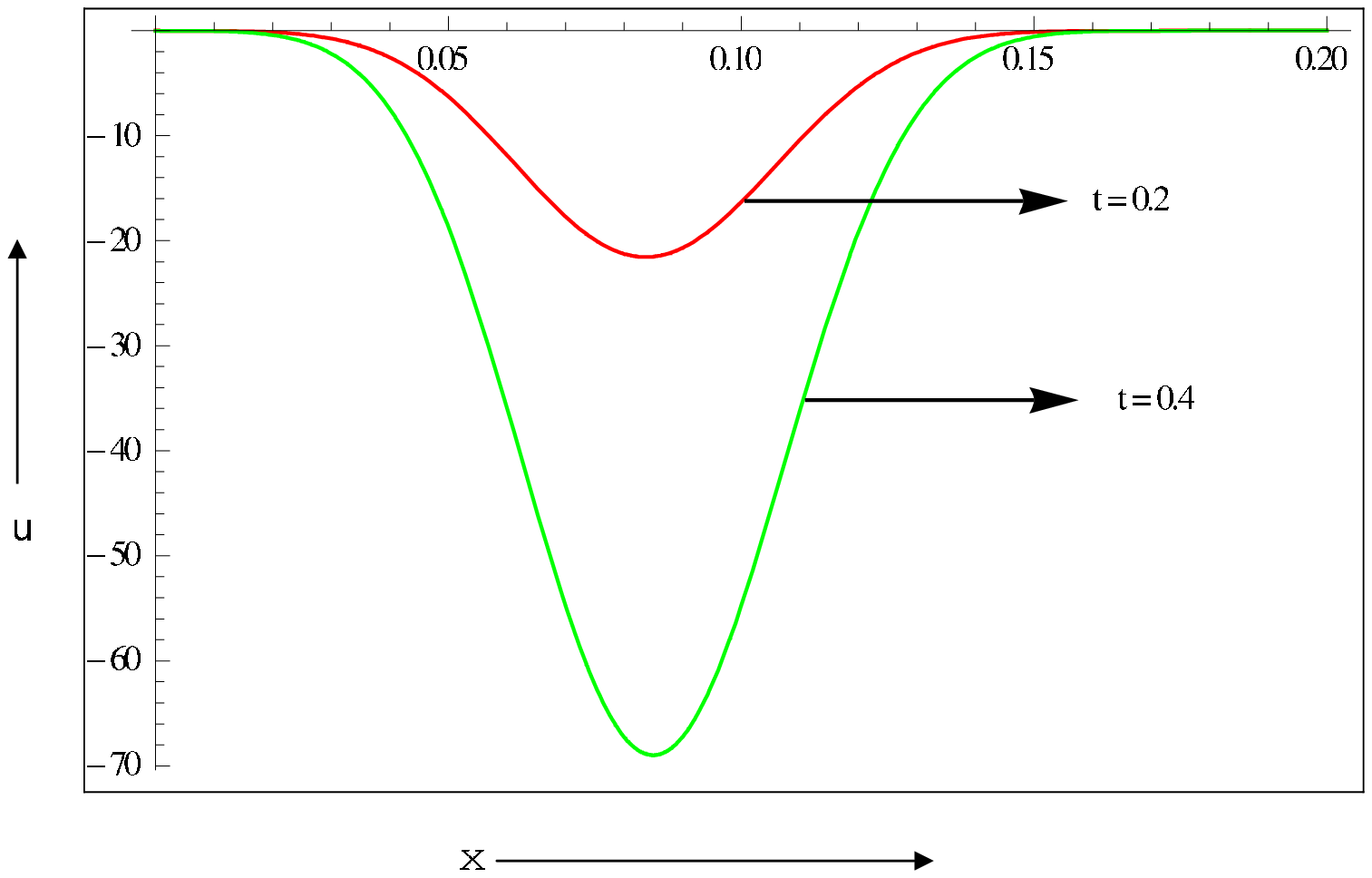}\\
Fig.1~~Distribution of displacement($u$) vs. $x$.
\vspace*{3.5cm}
$$  $$
\includegraphics[width=4.5in,height=2.5in ]{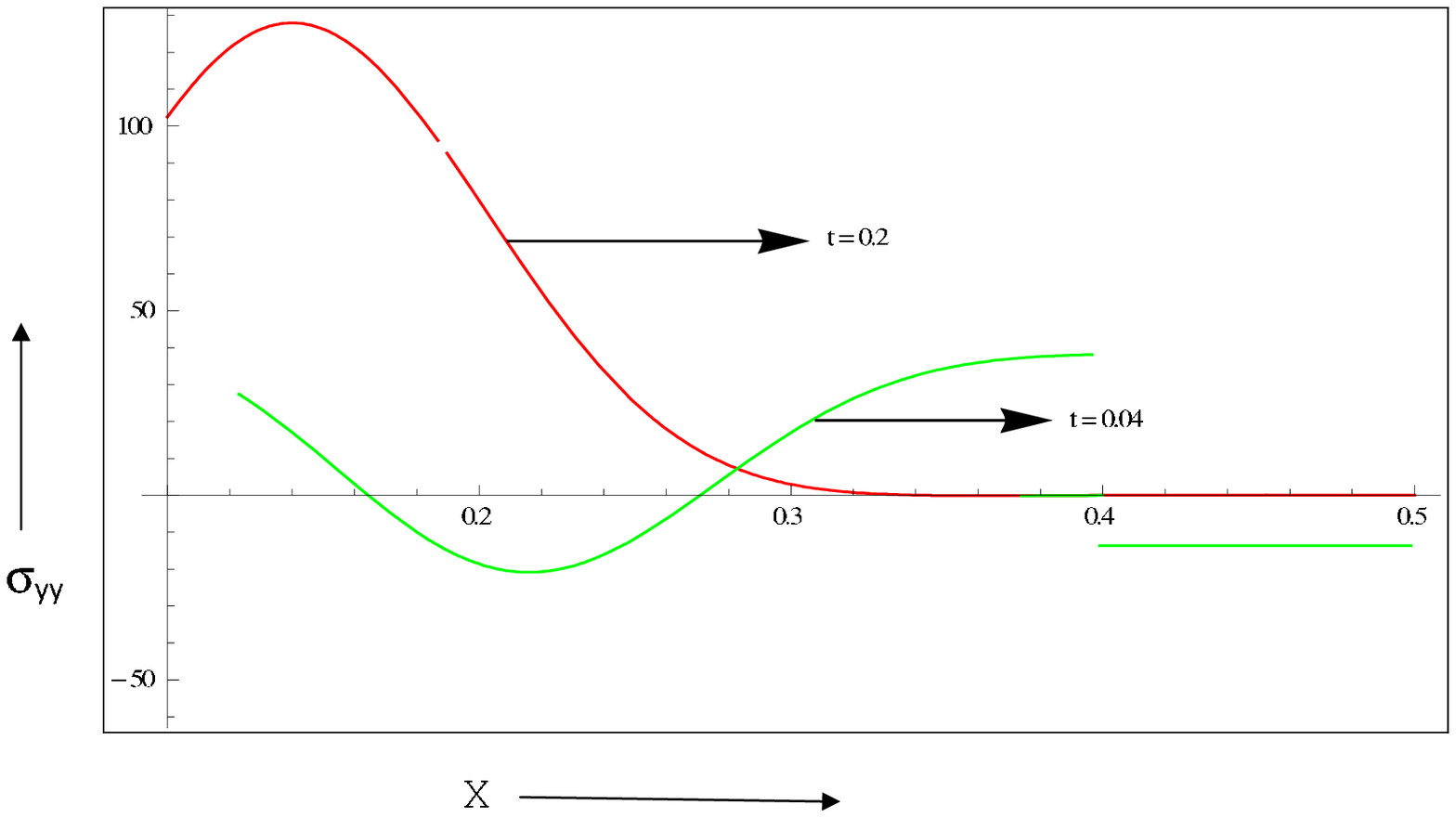}~~~~~~~~~\\
Fig.2~~Distribution of stress($\sigma_{yy}$) vs. $x$.
\end{minipage}\vspace*{.5cm}\\
\textheight 22.0cm

\textheight 22.0cm 
\begin{minipage}{1.0\textwidth}
$$  $$ $$  $$
 \includegraphics[width=4.0in,height=2.5in ]{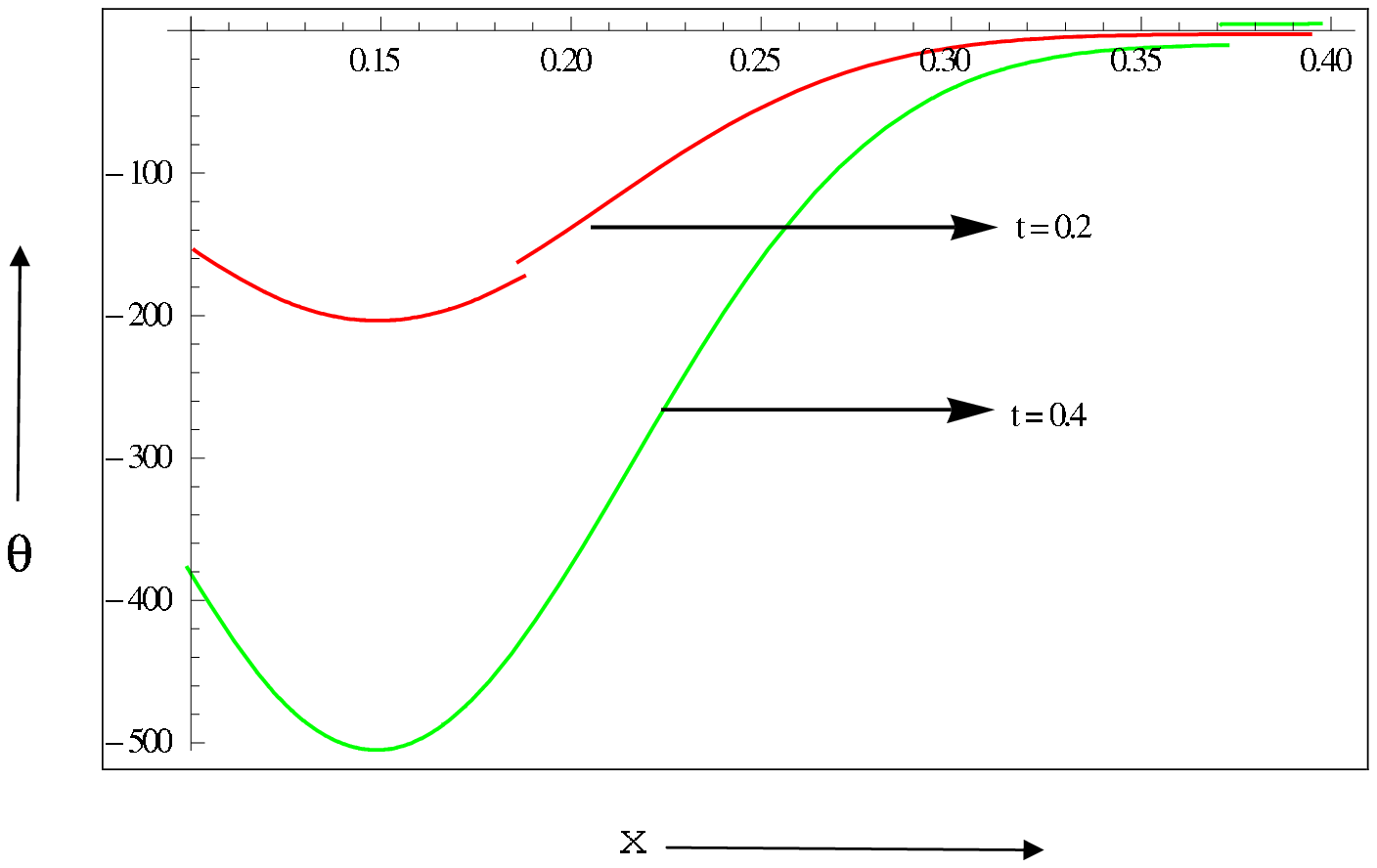}\\
Fig.3~~Distribution of temperature($\theta$) vs. $x$. \\
\end{minipage}

2.~~Figs.1-3 exhibit the variation of displacement, stresses,
temperature and induced magnetic field verses $x$ for fixed values of $t$ and rotation$(\Omega) = 4.95036\times10^{9}$. From Fig.1, it is clear that for fixed time$(t)$, the displacement component $u$ is continuous for all values of $x$ and the absolute value gradually increases in the region $0\leq x\leq 0.06$ and after that it gradually decreases as $x$ increases. From Fig.2, we see that the stress component $\sigma_{yy}$ is extensive in the region $0\leq x\leq 0.3$. Its value is gradually decreased as $x$ increases. Finally, it vanishes for the value of time $t_2$. $\sigma_{yy}$ is also compressive in the region $0.16\leq x\leq 0.27$ and $ x\geq 0.4$ otherwise it is extensive for the value of time $t_1$. 
       Fig.3 shows that the absolute value of temperature $(\theta)$ is decreased from $t_3$ to $t_2$ as $x$ increases. \\
\vspace*{4.0cm}
\textheight 22.0cm \pagebreak
\begin{minipage}{1.0\textwidth}
$$  $$
 \includegraphics[width=4.5in,height=2.5in ]{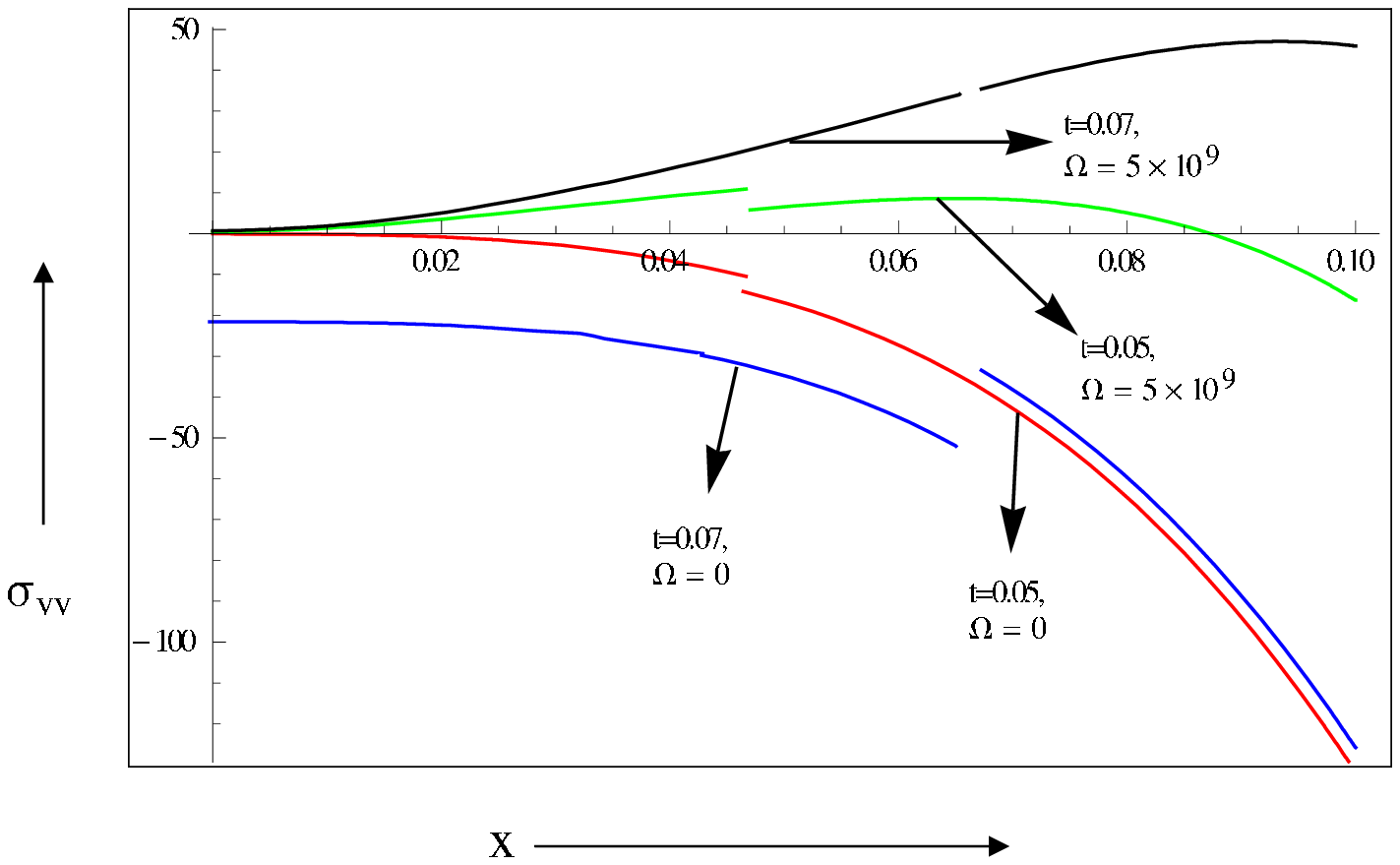}\\
Fig.4~~Distribution of stress($\sigma_{yy}$) vs. $x$ for fixed values of $t$ and $\Omega$ \\
\end{minipage}\vspace*{4.0cm}\\

\textheight 22.0cm 
\begin{minipage}{1.0\textwidth}
  \includegraphics[width=4.5in,height=2.5in ]{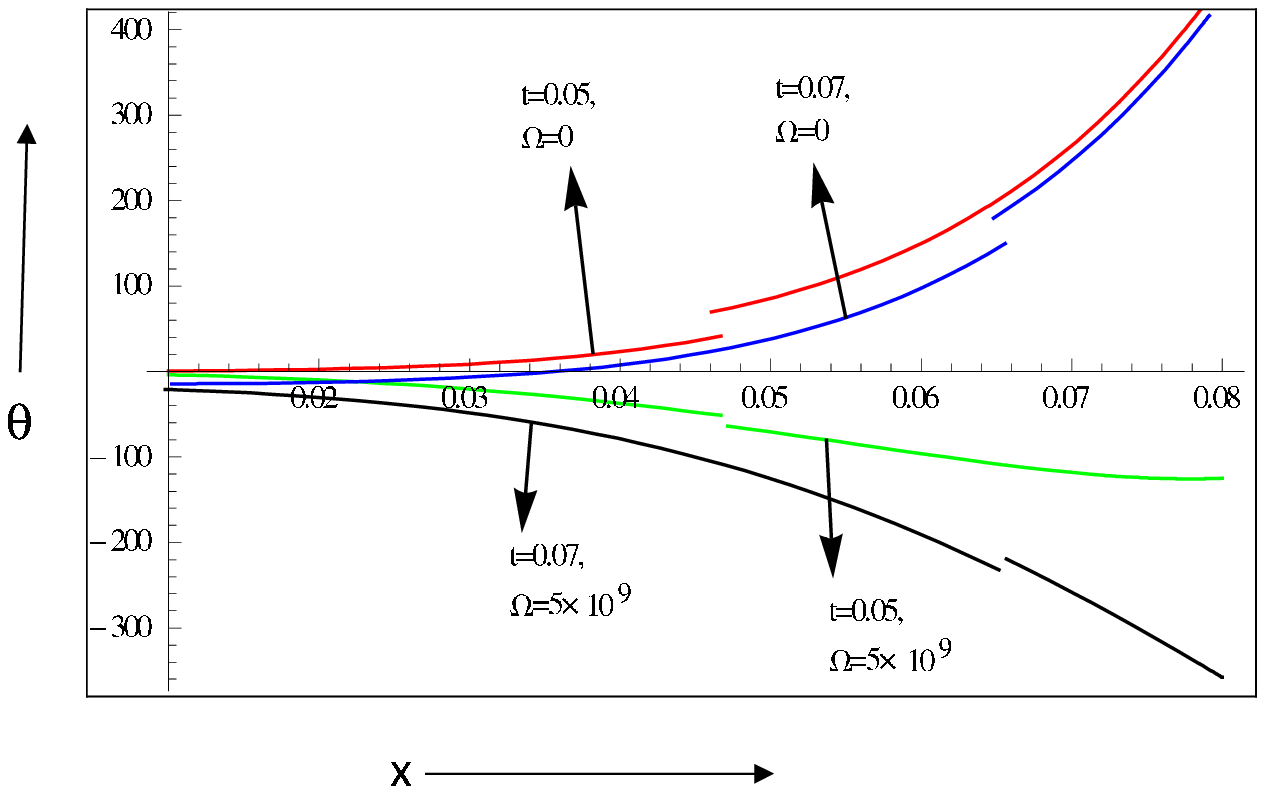}\\
Fig.5 ~~Distribution of temperature($\theta$) vs. $x$ for fixed values of $t$ and $\Omega$\\
\end{minipage}\vspace*{.5cm}\\

3.~~Figs.4-5 exhibit the variation of stress$(\sigma_{yy})$ and temperature$(\theta)$ verses $x$ for fixed values of 4 $t$ and $\Omega$.
       Fig.5 illustrates that for fixed time$(t)$ and $x$, the absolute value of stress component$(\sigma_{yy})$ gradually decreases as rotational vector $(\Omega)$ increases from $0$ to $5\times 10^{9}$. We also see that it is extensive for rotational vector $\Omega=5\times 10^{9}$ and also it becomes compressive when the rotation is withdrawn. Fig.6 also clearly shows that the value of temperature$(\theta)$ increases as rotational vector $(\Omega)$ decreases from $5\times 10^{9}$ to $0$. We also see that the absolute value of $\theta$ increases as $x$ increases. It is also clear that the effect of rotation changes the nature of temperature($\theta$).\\

{\bf{\underline{{Conclusions}}}}\\
The expressions for the thermoelastic displacement, temperature distribution, induced magnetic and electric field distributions also thermomechanical stresses for an isotropic half-space with rotational vector have been studied with the help of Lord-Shulman theory. The rotational vector$(\Omega)$ plays an important role in the expressions for the temperature distribution and thermomechanical stresses. For small time solutions, we have observed that the displacement $u(x,t)$ is continuous for all values of $x$ and $t$, i.e., it follows the continuum hypothesis. Also, each part of the solutions of displacement and stress components, temperature distribution, induced magnetic and electric field distributions are made up of two parts and that each part corresponds to a wave propagation with a finite speed. The first part of the each solutions involving the term $H(t-b_{10}x)$ represents 'elastic wave(e-wave)' which is travelling with also a finite speed of wave propagation $V_e=\frac{1}{b_{10}}$ at the wave front $\chi=\chi_1=\frac{t}{b_{10}}$ characterized by an exponential attenuation which is also influenced by $b_{11}$. The second part of the each solutions involving the term $H(t-b_{20}x)$ represents 'thermal wave($\theta$-wave)' which is travelling with also a finite speed of wave propagation $V_{\theta}=\frac{1}{b_{20}}$ at the wave front $\chi=\chi_2=\frac{t}{b_{20}}$ characterized by an exponential attenuation influenced by $b_{21}$.\\

\large{\bf{Appendix-I}}\\
$ L_{21}=\left[\begin{array}{clcr}
 C_{31} & 0 \\
 0 & C_{42}
  \end{array} \right]$, $ L_{22}=\left[\begin{array}{clcr}
 0 & C_{34} \\
 C_{43} & 0
  \end{array} \right]$,
$C_{31}=\frac{\alpha_0p^2}{\beta^2}-\frac{\varepsilon_1}{\beta^2}=a_1 p^2-a_2~;~C_{34}=\frac{\beta_0^2}{\beta^2}=a_3~;~
C_{42}= p+\tau_0 p^2~;~C_{43}= \varepsilon_2(p+\tau_0 p^2)$;\\
$a_{10}=\frac{1}{2}[\tau_0(a_3\varepsilon_2+1)+a_1+A]$;\\
$a_{11}=\frac{1}{2}[(a_3\varepsilon_2+1)+\frac{\{\tau_0(a_3\varepsilon_2+1)+a_1\}(a_3\varepsilon_2+1)-2a_1}{A}]$;\\
$a_{12}=\frac{1}{4}[-2a_2-\frac{[-2a_2\{\tau_0(a_3\varepsilon_2+1)+a_1\}+(a_3\varepsilon_2+1)^2+4a_2\tau_0]
[\{\tau_0(a_3\varepsilon_2+1)+a_1\}(a_3\varepsilon_2+1)-2a_1]}{A^3}]$;\\
$a_{20}=\frac{1}{2}[\tau_0(a_3\varepsilon_2+1)+a_1-A]$;\\
$a_{21}=\frac{1}{2}[(a_3\varepsilon_2+1)-\frac{\{\tau_0(a_3\varepsilon_2+1)+a_1\}(a_3\varepsilon_2+1)-2a_1}{A}]$;\\
$a_{22}=\frac{1}{4}[-2a_2+\frac{[-2a_2\{\tau_0(a_3\varepsilon_2+1)+a_1\}+(a_3\varepsilon_2+1)^2+4a_2\tau_0]
[\{\tau_0(a_3\varepsilon_2+1)+a_1\}(a_3\varepsilon_2+1)-2a_1]}{A^3}]$;\\
$A=[\{\tau_0(a_3\varepsilon_2+1)+a_1\}^2-4a_1\tau_0]^{\frac{1}{2}}$;\\
\large{\bf{Appendix-II}}\\
$L^{-1}[\bar{f}_1(p)\bar{f}_2(p)]={\int^{t}_{0}f_1(t-z)f_2(z)dt}$;\\
$L^{-1}(e^{-qp})=\delta(t-q)$;\\
$L^{-1}(p^{-\nu-1})=\frac{t^{\nu}}{\Gamma(\nu+1)}~~;Re~\nu>-1$;\\
$L^{-1}(p^{-\nu-1}e^{-\frac{q}{p}})=(\frac{t}{q})^{\frac{\nu}{2}}J_\nu(2\sqrt{qt})~~;Re~\nu>-1~,q>0$;\\
$L^{-1}(p^{-\nu-1}e^{\frac{q}{p}})=(\frac{t}{q})^{\frac{\nu}{2}}I_\nu(2\sqrt{qt})~~;Re~\nu>-1~,q>0$;\\
\large{\bf{Appendix-III}}\\
$L_1=b_{10}^2-b_{20}^2~~;$$~~L_2=2\{b_{10}b_{11}-b_{20}b_{21}\}~~;$\\
$L_{11}=a_3L_1b_{20}^2-L_1b_{20}^2+a_1L_1~~;$\\
$L_{12}=2a_3L_1b_{20}b_{21}-2L_1b_{20}b_{21}-a_3L_2b_{20}^2+L_2b_{20}^2-a_1L_2~~;$\\
$L_{21}=a_3L_1b_{10}^2-L_1b_{10}^2+a_1L_1~~;$\\
$L_{22}=2a_3L_1b_{10}b_{11}-2L_1b_{10}b_{11}-a_3L_2b_{10}^2+L_2b_{10}^2-a_1L_2~~;$\\
$b_1=L_{11}b_{10}~~;$ $~~b_2=L_{12}b_{10}+L_{11}b_{11}~~;$ $b_3=L_{12}b_{11}
+L_{11}b_{12}~~;$\\
$c_1=L_{21}b_{20}~~;$ $~~c_2=L_{22}b_{20}+L_{21}b_{21}~~;$ $c_3=L_{22}b_{21}
+L_{21}b_{22}~~; $\\
$z_1=2\sqrt{(t-xb_{10})xb_{12}}~~;$ $z_2=2\sqrt{-(t-xb_{20})xb_{22}}~~$\\
$d_0=L_{11}b_{10}^2-a_1L_{11}~~;$ $~~d_1=2L_{11}b_{10}b_{11}+L_{12}b_{10}^2-a_1L_{12}~~;$\\
$d_2=L_{11}b_{11}^2+2L_{11}b_{10}b_{12}+2L_{12}b_{10}b_{11}+a_2L_{11}~~;$\\
$d_3=2L_{11}b_{11}b_{12}+L_{12}b_{11}^2+2L_{12}b_{10}b_{12}+a_2L_{12}~~;$\\
$e_0=L_{21}b_{20}^2-a_1L_{21}~~;$ $~~e_1=2L_{21}b_{20}b_{21}+L_{22}b_{20}^2-a_1L_{22}~~;$\\
$e_2=L_{21}b_{21}^2+2L_{21}b_{20}b_{22}+2L_{22}b_{20}b_{21}+a_2L_{21}~~;$\\
$e_3=2L_{21}b_{21}b_{22}+L_{22}b_{21}^2+2L_{22}b_{20}b_{22}+a_2L_{22}$\\
$f_0=L_{11}b_{10}^2~~;$ $~~f_1=2L_{11}b_{10}b_{11}+L_{12}b_{10}^2~~;$\\
$f_2=L_{11}b_{11}^2+2L_{11}b_{10}b_{12}+2L_{12}b_{10}b_{11}~~;$\\
$f_3=2L_{11}b_{11}b_{12}+L_{12}b_{11}^2+2L_{12}b_{10}b_{12}~~;$\\
$g_0=L_{21}b_{20}^2~~;$ $~~g_1=2L_{21}b_{20}b_{21}+L_{22}b_{20}^2~~;$\\
$g_2=L_{21}b_{21}^2+2L_{21}b_{20}b_{22}+2L_{22}b_{20}b_{21}~~;$\\
$g_3=2L_{21}b_{21}b_{22}+L_{22}b_{21}^2+2L_{22}b_{20}b_{22}$\\
$h_0=L_{11}b_{10}~~;$ $~~h_1=L_{11}b_{11}+L_{12}b_{10}~~;$
$h_2=L_{11}b_{12}+L_{12}b_{11}~~;$
$h_3=L_{12}b_{12}~~;$\\
$k_0=L_{21}b_{20}~~;$ $~~k_1=L_{21}b_{21}+L_{22}b_{20}~~;$
$k_2=L_{21}b_{22}+L_{22}b_{21}~~;$
$k_3=L_{22}b_{22}$\\
$m_0=(a_3-1)L_{11}b_{10}^2+a_1L_{11}~~;$ $~~m_1=(a_3-1)(L_{12}b_{10}^2+2L_{11}b_{10}b_{11})+a_1L_{12}~~;$\\
$m_2=(a_3-1)(2L_{12}b_{10}b_{11}+L_{11}b_{11}^2+2L_{11}b_{10}b_{12})-a_2L_{11}~~;$\\
$m_3=(a_3-1)(L_{12}b_{11}^2+2L_{12}b_{10}b_{12}+2L_{11}b_{11}b_{12})-a_2L_{12}~~;$\\
$n_0=(a_3-1)L_{21}b_{20}^2+a_1L_{21}~~;$ $~~n_1=(a_3-1)(L_{22}b_{20}^2+2L_{21}b_{20}b_{21})+a_1L_{22}~~;$\\
$n_2=(a_3-1)(2L_{22}b_{20}b_{21}+L_{21}b_{21}^2+2L_{21}b_{20}b_{22})-a_2L_{21}~~;$\\
$n_3=(a_3-1)(L_{22}b_{21}^2+2L_{22}b_{20}b_{22}+2L_{21}b_{21}b_{22})-a_2L_{22}~~;$\\
$r_0=\{(1-\frac{2}{\beta_0^2})a_3-1\}L_{11}b_{10}^2+a_1L_{11}~~;$ $~~r_1=\{(1-\frac{2}{\beta_0^2})a_3-1\}(L_{12}b_{10}^2+2L_{11}b_{10}b_{11})+a_1L_{12}~~;$\\
$r_2=\{(1-\frac{2}{\beta_0^2})a_3-1\}(2L_{12}b_{10}b_{11}+L_{11}b_{11}^2+2L_{11}b_{10}b_{12})-a_2L_{11}~~;$\\
$r_3=\{(1-\frac{2}{\beta_0^2})a_3-1\}(L_{12}b_{11}^2+2L_{12}b_{10}b_{12}+2L_{11}b_{11}b_{12})-a_2L_{12}~~;$\\
$s_0=\{(1-\frac{2}{\beta_0^2})a_3-1\}L_{21}b_{20}^2+a_1L_{21}~~;$ $~~s_1=\{(1-\frac{2}{\beta_0^2})a_3-1\}(L_{22}b_{20}^2+2L_{21}b_{20}b_{21})+a_1L_{22}~~;$\\
$s_2=\{(1-\frac{2}{\beta_0^2})a_3-1\}(2L_{22}b_{20}b_{21}+L_{21}b_{21}^2+2L_{21}b_{20}b_{22})-a_2L_{21}~~;$\\
$s_3=\{(1-\frac{2}{\beta_0^2})a_3-1\}(L_{22}b_{21}^2+2L_{22}b_{20}b_{22}+2L_{21}b_{21}b_{22})-a_2L_{22}~~;$\\

\end{document}